\documentclass{moriond}

\def\Journal#1#2#3#4{{#1} {\bf #2}, #3 (#4)}

\def\PRL{\em Phys. Rev. Lett.}
\def\PRD{{\em Phys. Rev.} D}

\def\JHEP{{\em JHEP}}
\def\EPJC{{\em Eur. Phys. J.} C}
\def\JINST{\em JINST}

\def\be{\begin{equation}}
\def\ee{\end{equation}}
\def\bea{\begin{eqnarray}}
\def\eea{\end{eqnarray}}

\usepackage{xspace}

\begin{document}
\vspace*{2cm}
\title{Rare top quark production and decays at ATLAS and CMS}

\author{ K. Skovpen (on behalf of the ATLAS and CMS Collaborations) }

\address{IIHE - VUB, Pleinlaan 2, 1050 Brussels, Belgium}

\maketitle\abstracts{The most recent studies in the top quark
sector are reviewed with the focus on the rare production mechanisms
and suppressed decays. The experimental results obtained with the ATLAS and CMS
detectors in proton-proton collisions at the center-of-mass energy of
13 TeV include the measurements of the associated production of top quark pairs with vector bosons
(${\rm t\bar{t}W}$, ${\rm t\bar{t}Z}$, ${\rm t\bar{t}\gamma}$), the first evidence for the ${\rm
t(\bar{t})\gamma q}$ process, the first observation of the ${\rm t(\bar{t})Zq}$ production,
the study of the ${\rm t\bar{t}+b\bar{b}}$ and ${\rm t\bar{t}+t\bar{t}}$
processes, as well as searches for lepton flavour violation in top
quark decays and effective field theory interpretations. The
experimental results show good agreement with the theoretical
predictions.}

{\let\thefootnote\relax\footnote{{\textcopyright\xspace 2019 CERN for the benefit
of the ATLAS and CMS Collaborations. CC-BY-4.0 license.}}}

\section{Introduction}

The top quark takes an important place in the standard model (SM).
It's mass and distinctive experimental decay signature in the experiment
makes it possible to study a number of very rare production mechanisms, 
as well as extremely rare decay modes. Top quarks produced in pairs, 
as well as singly produced particles, were already observed at the LHC.
The rare production of top quarks with vector bosons and additional
quarks is associated with small cross sections and is also challenging
due to complex final states. Study of these rare processes allows us
to probe interactions of the top quark with other SM particles and to
search for possible anomalous phenomena. 

\section{Study of associated production of top quark pairs with vector bosons}

The study of the top quark pair production (${\rm t\bar{t}}$) in association with a W or Z boson is
important because these topologies can receive sizeable contributions
from new physics, and, in addition, the ${\rm t\bar{t}Z}$ production represents the main
channel to directly measure the top quark couplings to the Z boson.
Moreover, the precise measurements of the production cross sections of
these processes are essential for the study of the ${\rm t\bar{t}H}$
production where the ${\rm t\bar{t}W}$ and ${\rm t\bar{t}Z}$ events represent one of the dominant backgrounds in the multilepton
analysis channels. 

The study of the ${\rm t\bar{t}W}$ production at ATLAS~\cite{ATLASdet} is done in
the dilepton same-sign and trilepton channels, while the ${\rm
t\bar{t}Z}$ process
is looked for in the dilepton opposite-sign, trilepton and four-lepton
final states~\cite{ttV_ATLAS_1}. In the analysis of the ${\rm t\bar{t}Z}$
process using dilepton opposite-sign final states the prompt lepton background originates from
${\rm Z}$+jets and ${\rm t\bar{t}}$ events, while these processes represent a non-prompt
lepton background in the other channels, also for the case of ${\rm
t\bar{t}W}$. The
prompt lepton background is additionally associated with the diboson
production and is one of the dominant backgrounds along with the
non-prompt leptons. The analysis proceeds with defining control and
signal regions with multiple exclusive event categories based
on the number of leptons split into different flavour and sign with an
additional selection based on the number of jets and b-tagged jets. 

\begin{figure}[hbtp]
\begin{center}
\includegraphics[width=0.65\linewidth]{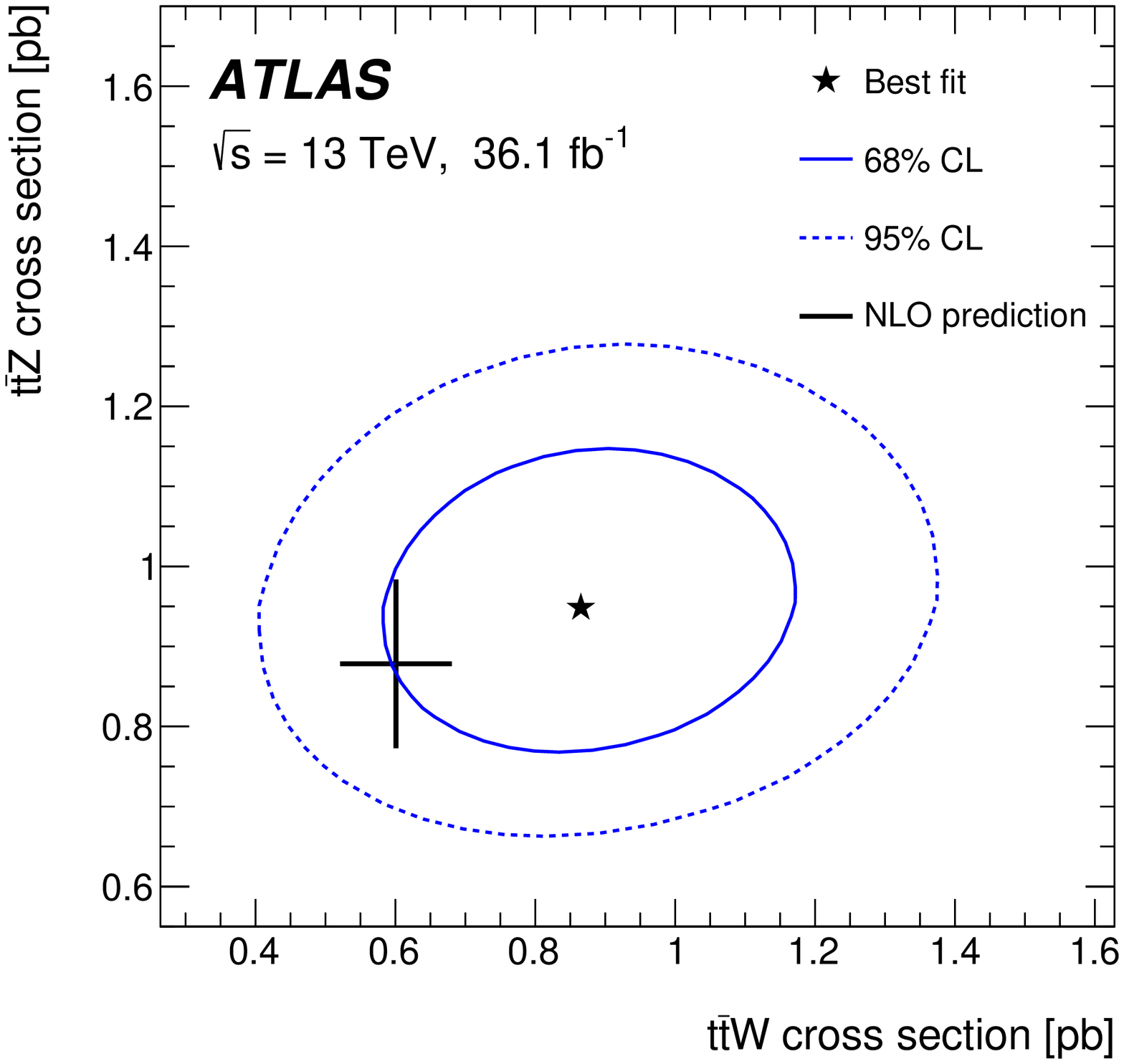}
\caption{The results of the simultaneous fit to the ${\rm t\bar{t}Z}$ and
${\rm t\bar{t}W}$ cross sections with the 68\% and 95\% confidence level
contours compared to the NLO theoretical predictions~\protect\cite{ttV_ATLAS_1}.}
\label{fig:ttV}
\end{center}
\end{figure}

The analysis of the ${\rm t\bar{t}Z}$ production allows to perform an
effective field theory (EFT) study of anomalous contributions to the
${\rm t\bar{t}Z}$ vertex with obtaining
constraints on the Wilson coefficients of the respective dimension-six
operators. In such interpretation the ${\rm t\bar{t}Z}$ event rate can be expressed
as a quadratic function of the Wilson coefficients where the linear
terms results from the interference between beyond the SM (BSM) and SM operators.
The fits to the measured distributions are done to obtain the EFT
constrains in the case when both the quadratic and linear terms are
kept, as well as when the quadratic terms are omitted. The obtained
constraints represent competitive results to the existing direct
and indirect limits. The inclusive ${\rm t\bar{t}Z}$ and ${\rm
t\bar{t}W}$ cross sections are
measured with $\simeq$~10\% and $\simeq$~20\% precision, respectively, and are comparable
to next-to-leading order (NLO) theoretical uncertainties, as shown in Fig.~\ref{fig:ttV}.

The analysis at CMS~\cite{CMSdet} studies final states with two same-sign
leptons for the ${\rm t\bar{t}W}$ process, while the ${\rm t\bar{t}Z}$ production is looked for in the
trilepton and four lepton final states~\cite{ttV_CMS_1}. The prompt and non-prompt
lepton backgrounds are validated in control regions in data. The
analysis uses an improved multivariate-analysis-based (MVA) lepton identification with respect
to the previous iterations of these studies. The ${\rm t\bar{t}W}$ and
${\rm t\bar{t}Z}$ cross sections are extracted from combined fit over several exclusive event
categories defined by the final MVA-based discriminant and the total
number of jets and b-tagged jets. The measured inclusive cross
sections show good agreement with NLO predictions.

\begin{figure}[hbtp]
\begin{center}
\includegraphics[width=0.43\linewidth]{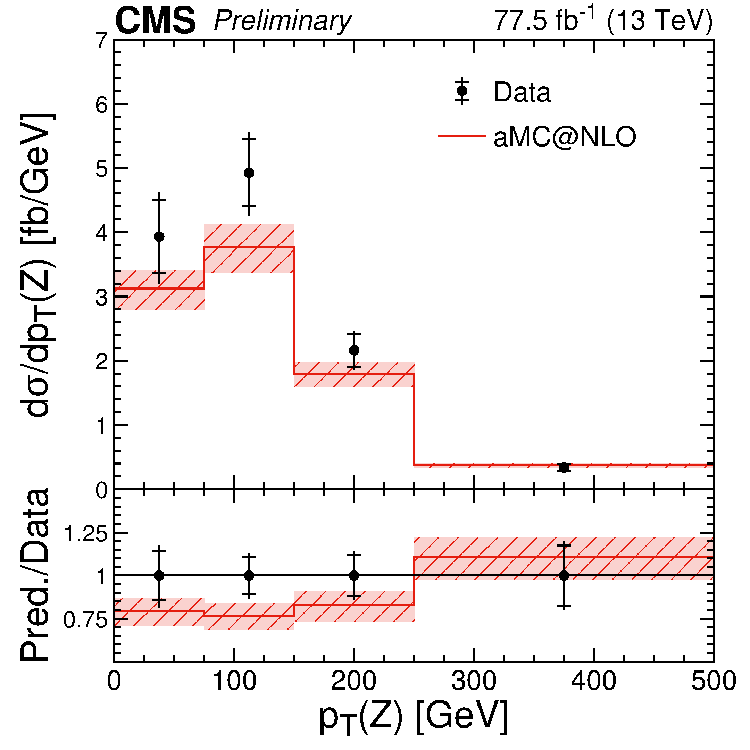}
\includegraphics[width=0.43\linewidth]{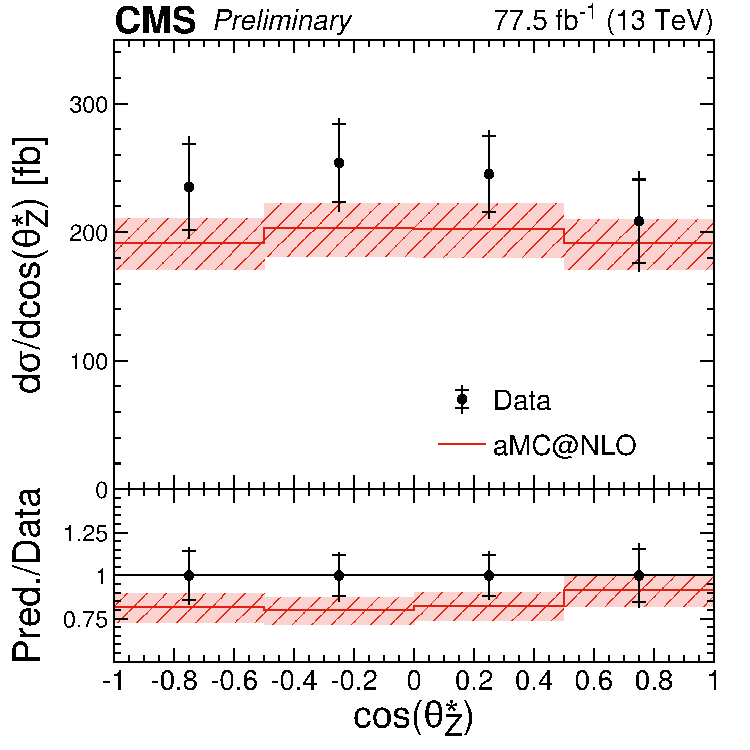}
\caption{Comparison between data and MC prediction for differential ${\rm t\bar{t}Z}$ cross sections as a
function of the transverse momentum of the Z boson (left) and the cosine
of the angle between the Z boson and the negatively charged lepton
from the Z boson decay in the Z boson rest frame
(right)~\protect\cite{ttZ_CMS_2}. The hatched band includes the theory uncertainties in the
prediction.}
\label{fig:ttZ}
\end{center}
\end{figure}

By including more data it becomes possible to probe differential cross
sections of the ${\rm t\bar{t}Z}$ production. The study done at CMS
measures differential distributions of the ${\rm t\bar{t}Z}$ cross
section using kinematic variables sensitive to ${\rm t-Z}$ anomalous
interactions~\cite{ttZ_CMS_2}. The measured cross sections are interpreted in two
frameworks. The first approach uses an anomalous-coupling Lagrangian
based on the neutral vector and axial vector current couplings, as
well as the weak magnetic and electric dipole interaction couplings.
The second interpretation is EFT-based which considers four
dimension-six operators which induce electroweak dipole moments and
anomalous neutral-current interactions. The ${\rm t\bar{t}Z}$ inclusive cross
section in this recent analysis is now measured with an improved
precision of $\simeq$~10\%. Some measured differential cross
sections are presented in Fig.~\ref{fig:ttZ}.

The analysis of the ${\rm t\bar{t}\gamma}$ production represents an important
study of the top-photon electroweak couplings  where the kinematic
distributions of the radiated photon, such as transverse momentum, are
especially sensitive to new physics contributions. The measurement of the
${\rm t\bar{t}\gamma}$ differential cross section also provides an important
information on the ${\rm t\bar{t}}$ spin correlations and charge asymmetry, and
is complementary to the other ${\rm t\bar{t}}$ measurements. The
${\rm t\bar{t}\gamma}$ process is studied at ATLAS in the channels with one or two leptons and the
results of the differential measurements are compared to leading-order
and NLO predictions~\cite{ttg_ATLAS_1}. The previously observed disagreements in the large values
in the distribution of the azimuthal angular difference between the
two leptons are now significantly mitigated with moving to the NLO
event generation. 

\section{Study of single top quark associated production with vector bosons}

The associated production of a top quark with a photon (${\rm
t(\bar{t})\gamma q}$) is an important
process which is sensitive to the charge, as well as the electric and
magnetic moments of the top quark. The search for this process is done
at CMS in the t-channel considering the final state with one muon, one
photon, one b-tagged jet and one forward jet~\cite{tgq_CMS_1}. 

\begin{figure}[hbtp]
\begin{center}
\includegraphics[width=0.5\linewidth]{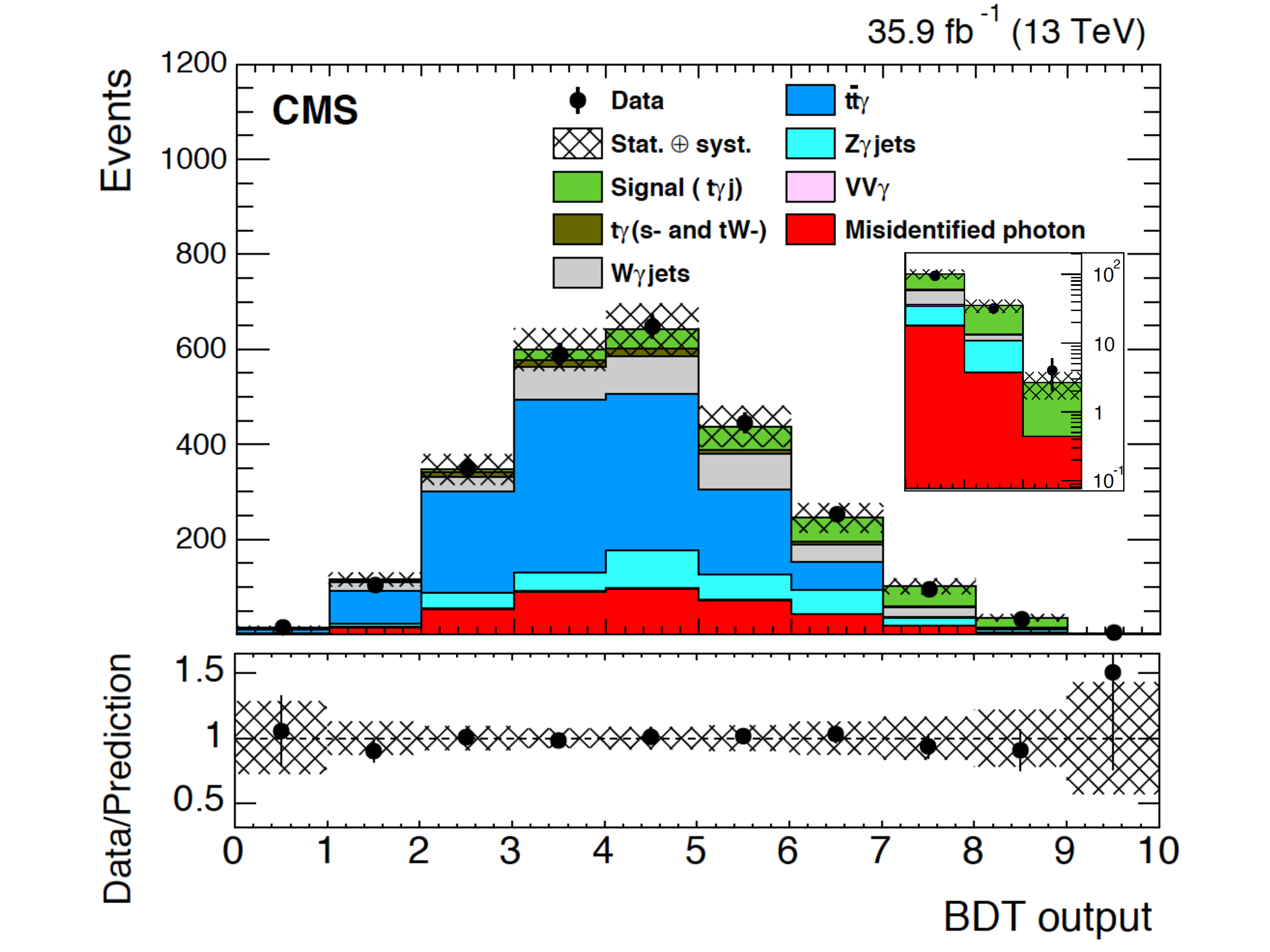}
\caption{The boosted decision tree output distribution for data and SM predictions after
the fit in the analysis of the ${\rm t(\bar{t})\gamma q}$
process~\protect\cite{tgq_CMS_1}.}
\label{fig:tgq}
\end{center}
\end{figure}

\noindent The presence of the
forward light flavour energetic jet is a very characteristic signature
of the single top quark associated production with vector bosons. The
dominant background includes the ${\rm t\bar{t}\gamma}$ production, among other
contributions. The analysis uses a boosted decision tree-based
discriminator to suppress various backgrounds, as shown in
Fig.~\ref{fig:tgq}. This study resulted in the first evidence for this
process at 4.4~(3.0)~$\sigma$ observed (expected).

\begin{figure}[hbtp]
\begin{center}
\includegraphics[width=0.4\linewidth]{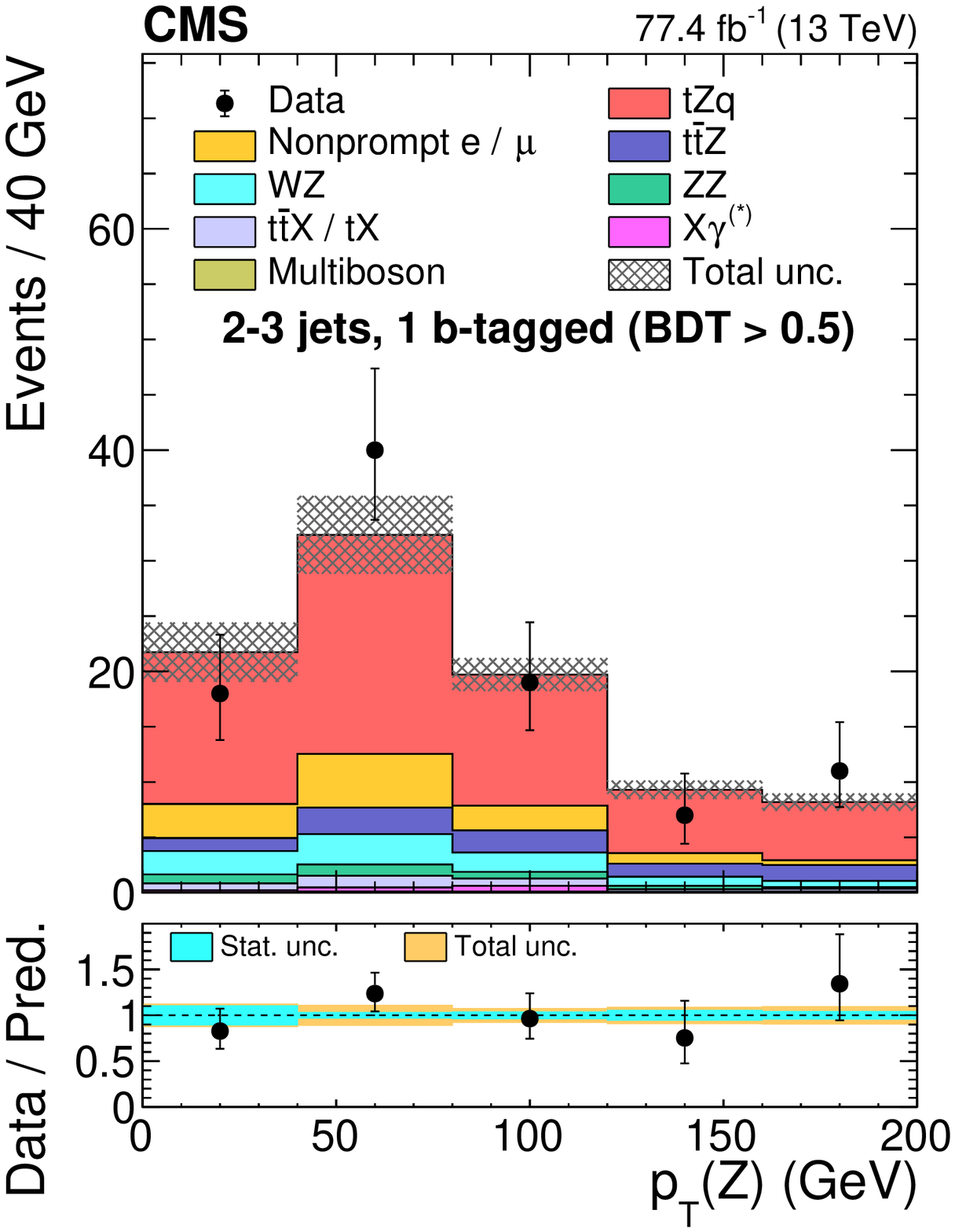}
\includegraphics[width=0.4\linewidth]{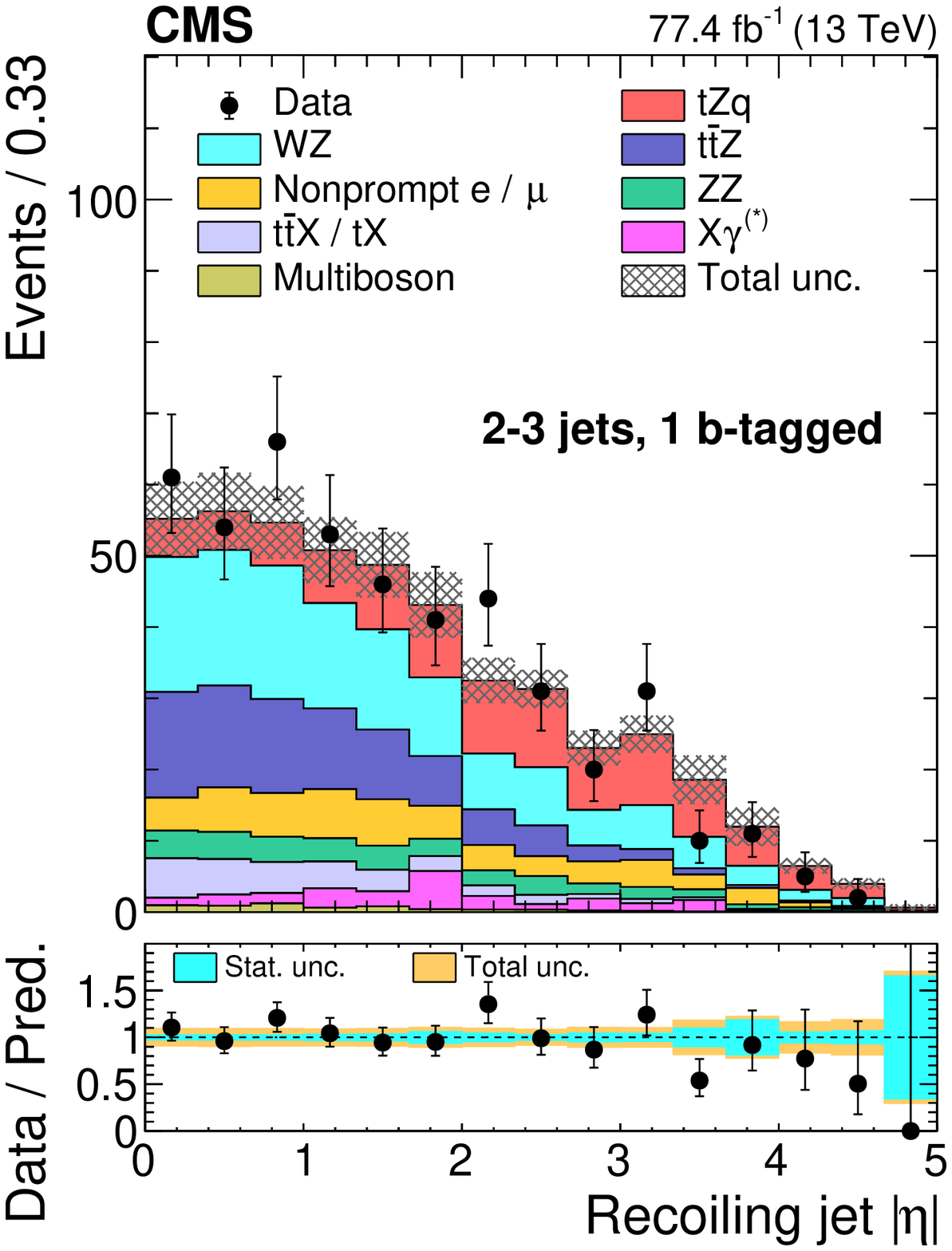}
\caption{Comparison between the number of expected and observed events
for the distributions of the recoiling jet $|\eta|$ (left) and the
transverse momentum of the ${\rm Z}$ boson after the additional selection on
the final discriminant (right)~\protect\cite{tZq_CMS_1}.}
\label{fig:tZq}
\end{center}
\end{figure}

The production of a top quark in association with a Z boson (${\rm
t(\bar{t})Zq}$) is
sensitive to anomalous ${\rm WWZ}$ triple-gauge and ${\rm tZ}$ couplings. The analysis
of this production at CMS was done in the final state with three
leptons~\cite{tZq_CMS_1}. This study uses an
improved lepton identification which allowed to boost the final
sensitivity in this search. A simultaneous fit is performed over
several event categories to extract the signal with the sensitivity of
8.2~(7.7)~$\sigma$ observed (expected) leading to the first observation
of this process. Comparisons after the final event selection criteria 
between data and predictions are shown in Fig.~\ref{fig:tZq}.

\section{Study of top quark pair production in association with heavy
flavour quarks}

\begin{figure}[hbtp]
\begin{center}
\includegraphics[width=0.65\linewidth]{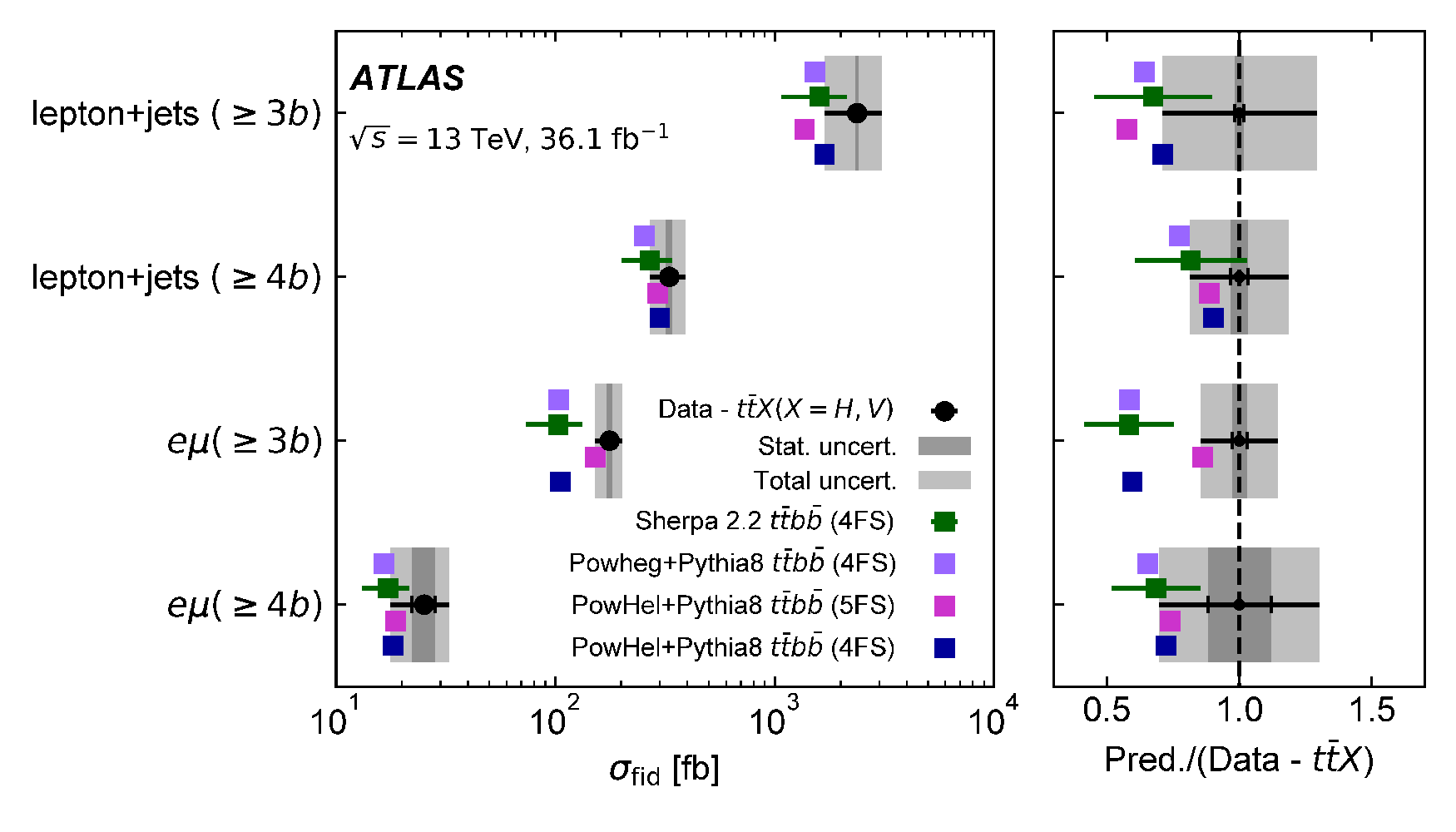}
\caption{The measured fiducial cross sections, with ${\rm t\bar{t}H}$ and
${\rm t\bar{t}V (V=W,Z,\gamma)}$  contributions subtracted from data, compared with
${\rm t\bar{t}+b\bar{b}}$ predictions~\protect\cite{ttbb_ATLAS_1}.}
\label{fig:ttbb}
\end{center}
\end{figure}

The production of ${\rm t\bar{t}}$ with additional jets is associated with
large theoretical uncertainties due to the presence of two different
scales of the top quark mass and the jet transverse momentum. The
measurements of the associated production of top quark pairs with b quarks 
(${\rm t\bar{t}+b\bar{b}}$) are done at ATLAS in
single lepton and dilepton final states~\cite{ttbb_ATLAS_1}.
This study is an important test of QCD predictions with providing a better
estimation of one of the main backgrounds in the ${\rm t\bar{t}H (H
\rightarrow b\bar{b})}$ analysis. The
${\rm t\bar{t}+b\bar{b}}$ component is extracted from data using MC templates defined by
the flavour of additional quark-jets. The measured inclusive fiducial
cross sections generally exceed the ${\rm t\bar{t}+b\bar{b}}$ NLO predictions but are still
compatible within the total uncertainties. The inclusive cross section
is measured with precision of $\simeq$~20\% and is better than in the
theoretical calculations. The measured cross sections in fidual
region is presented in Fig.~\ref{fig:ttbb}.

\begin{figure}[hbtp]
\begin{center}
\includegraphics[width=0.55\linewidth]{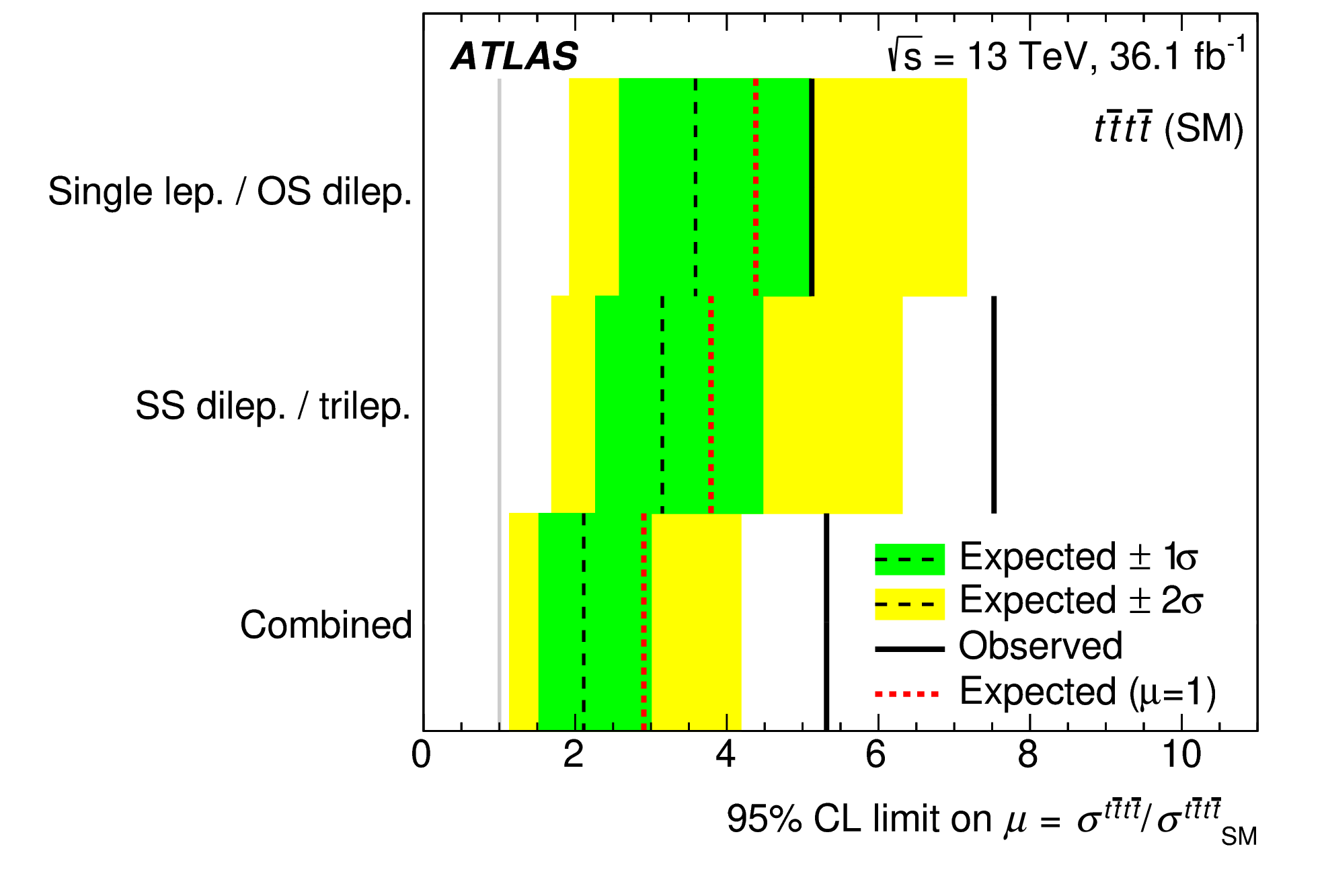}
\caption{Summary of the 95\% confidence level limits on the
${\rm t\bar{t}+t\bar{t}}$ production relative to the SM prediction in the individual
channels and for the combination~\protect\cite{tttt_ATLAS_1}.}
\label{fig:4top}
\end{center}
\end{figure}

Another rare process that allows to study the QCD predictions is the
four top quark production that is also sensitive to the top quark
Yukawa coupling. The search for this process at ATLAS is done in single lepton and dilepton
opposite-sign channels~\cite{tttt_ATLAS_1}. Events are categorised based on the number of
jets and b-tagged jets. There are several validation regions defined to
perform the measurement of the b tagging efficiencies adapted to the
topology of these events and to extrapolate it to the signal regions. The results
are combined with the previously published dilepton same-sign and
multilepton results. This combination has resulted in the exclusion of
the cross section of ${\rm t\bar{t}+t\bar{t}}$ production down to $\simeq$~$5\times$
the predicted value at 2.8~(1.0)~$\sigma$ observed (expected), with
the final limits presented in Fig.~\ref{fig:4top}. The analysis also includes an EFT interpretation
optimised for four-top contact interactions. 

The search for the ${\rm t\bar{t}+t\bar{t}}$ production in similar final states is also
done at CMS~\cite{tttt_CMS_1}. Several event categories are defined based on the number
of reconstructed jets and b-tagged jets used in a simultaneous fit to
extract the signal. A combination with dilepton same-sign and
trilepton results is also performed. The EFT interpretation was done
for four-fermion operators which contribute to the ${\rm
t\bar{t}+t\bar{t}}$ production.
The combined sensitivity to the ${\rm t\bar{t}+t\bar{t}}$ production reaches
1.4~(1.1)~$\sigma$ observed (expected).

\section{Effective field theory study in dilepton events}

The EFT interpretations are becoming an essential part of many analysis
studying top quarks. The full EFT at NLO interpretation is done for the top quark production in the dilepton
final state at CMS, which mainly includes top quarks produced in
${\rm t\bar{t}}$ and ${\rm t(\bar{t})W}$ processes~\cite{EFT_CMS_1}. There are several types of operators which
contribute to the production of these events, including operators
associated with the ${\rm Wtb}$ couplings, chromomagnetic dipole moment,
triple gluon field and flavour-changing neutral currents. The
interference between the ${\rm t\bar{t}}$ and ${\rm t(\bar{t})W}$
processes is removed. The EFT constrains are set through the fit of a
Neural Network discriminants trained to distinguish between the EFT contributions
and the SM prediction.

\section{Search for lepton-flavour violation}

In addition to the rare top quark production, one can also search for rare
decays of these particles. One such analysis is done at ATLAS to
search for the charged lepton-flavour violation (LFV) with the
model-independent approach in three-particle decays of top quarks~\cite{LFV_ATLAS_1}. 
The LFV decays of top quarks are extremely suppressed in the SM and 
any deviations from these zero rates would indicate the presence of new physics. 
The analysis is based on the study of the three-lepton final state to
set limits several EFT operators, including the axial-vector, scalar, pseudo-scalar and lepton-quark
interactions. The probability of observing LFV top quark decays is
excluded down to the $\simeq$~$10^{-5}$ level and this constraint is more
stringent than the current indirect
limits set at $\simeq$~$10^{-3}$.

\section{Conclusion}

The experiments done with the ATLAS and CMS detectors at the LHC
provide us with a great opportunity to study very rare
processes with top quarks. Recently, we have observed for the first
time the process with the production of single top quarks in association with a Z boson, as well as
have obtained the first evidence for the single top quark production
with a photon. The study of the underlying physics in these processes and searches for BSM phenomena 
will proceed with even more data in the coming years.

\end{document}